\documentstyle[11pt,aaspp4]{article}

\def\la{\mathrel{\hbox{\rlap{\hbox{\lower4pt\hbox{$\sim$}}}\hbox{$<$}}}}
\def\ga{\mathrel{\hbox{\rlap{\hbox{\lower4pt\hbox{$\sim$}}}\hbox{$>$}}}}

\slugcomment{Accepted to {\it The Astrophysical Journal}}

\begin{document}

\title{Evidence for a Molecular Cloud Origin for Gamma-Ray Bursts:  Implications for the Nature of Star Formation in the Universe}

\author{Daniel E. Reichart\altaffilmark{1,2} and Paul A. Price\altaffilmark{1,3}}

\altaffiltext{1}{Department of Astronomy, California Institute of Technology, Mail Code 105-24, 1201 East California Boulevard, Pasadena, CA 91125}
\altaffiltext{2}{Hubble Fellow}
\altaffiltext{3}{Research School of Astronomy and Astrophysics, Mount Stromlo Observatory, Cotter Road, Weston, ACT, 2611, Australia}

\begin{abstract}

It appears that the majority of rapidly-, well-localized gamma-ray bursts with undetected, or dark, optical afterglows, or `dark bursts' for short, occur in clouds of size $R \ga 10L_{49}^{1/2}$ pc and mass $M \ga 3\times10^5L_{49}$ M$_{\sun}$, where $L$ is the isotropic-equivalent peak luminosity of the optical flash.  We show that clouds of this size and mass cannot be modeled as a gas that is bound by pressure equilibrium with a warm or hot phase of the interstellar medium (i.e., a diffuse cloud):  Such a cloud would be unstable to gravitational collapse, resulting in the collapse and fragmentation of the cloud until a burst of star formation re-establishes pressure equilibrium within the fragments, and the fragments are bound by self-gravity (i.e., a molecular cloud).  Consequently, dark bursts probably occur in molecular clouds, in which case dark bursts are probably a byproduct of this burst of star formation if the molecular cloud formed recently, and/or the result of lingering or latter generation star formation if the molecular cloud formed some time ago.  We then show that if bursts occur in Galactic-like molecular clouds, the column densities of which might be universal, the number of dark bursts can be comparable to the number of bursts with detected optical afterglows:  This is what is observed, which suggests that the bursts with detected optical afterglows might also occur in molecular clouds.  We confirm this by modeling and constraining the distribution of column densities, measured from absorption of the X-ray afterglow, of the bursts with detected optical afterglows:  We find that this distribution is consistent with the expectation for bursts that occur in molecular clouds, and is not consistent with the expectation for bursts that occur in diffuse clouds.  Consequently, we find that all but perhaps a few bursts, dark or otherwise, probably occur in molecular clouds.  Finally, we show that the limited information that is available on the column densities of the dark bursts is not consistent with the idea that the dark bursts occur in the nuclear regions of ultraluminous infrared/submillimeter-bright galaxies, from which we draw conclusions about the nature of star formation in the universe.

\end{abstract}

\keywords{dust, extinction --- galaxies: starburst --- gamma rays: bursts --- infrared: galaxies --- ISM: clouds --- stars: formation}

\section{Introduction}

In Reichart \& Yost (2001), we show that about 60\% of all bursts, and about 80\% of rapidly-, well-localized gamma-ray bursts with undetected, or dark, optical afterglows, or `dark bursts' for short, have afterglows that are fainter than R $= 24$ mag 18 hours after the burst.  Furthermore, we show that, with the exception of perhaps a few bursts, the dark bursts are most likely the result of circumburst\footnote{By `circumburst', we mean within the circumburst cloud.} extinction, and the density of the circumburst medium probably spans many orders of magnitude, from densities possibly as low as densities that are typical of the Galactic disk to densities probably as high as densities that are typical of dense clouds.  

In Reichart (2001b; see also Waxman \& Draine 2000; Fruchter, Krolik \& Rhoads 2001), we show that sublimation of dust by the optical flash, and fragmentation of dust by the burst and afterglow, prevent clouds that are smaller than $R \sim 10L_{49}^{1/2}$ pc, where $L = 10^{49}L_{49}$ erg s$^{-1}$ is the 1 -- 7.5 eV isotropic-equivalent peak luminosity of the optical flash, from significantly extinguishing their afterglows.  Given the finding of Reichart \& Yost (2001) that circumburst extinction appears to be responsible for most of the dark bursts, we find in Reichart (2001b) that most of the dark bursts occur in clouds of size $R \ga 10L_{49}^{1/2}$ pc and column density $N_H \ga 5\times10^{21}$ cm$^{-2}$, and hence mass $M \ga 3\times10^5L_{49}$ M$_{\sun}$.  

In this paper, we consider the equilibrium properties of clouds, and show that clouds of these sizes and masses, and hence the circumburst clouds of dark bursts, cannot be diffuse, but rather are probably molecular, in which case they are active regions of star formation (\S 2).  In \S 3, we show that Galactic-like molecular clouds have sufficiently high column densities to make bursts dark at optical wavelengths, and that sublimation of dust by the optical flash and fragmentation of dust by the burst and afterglow can leave a comparable number of bursts -- bursts that are sufficiently near the earth-facing side of the cloud and/or in sufficiently small clouds -- relatively unextinguished at optical wavelengths, which is in agreement with what is observed.  In \S 4, we model and constrain the column density distribution of the bursts with detected optical afterglows, and show that these bursts are also consistent with a molecular cloud origin, and are not consistent with a diffuse cloud origin.  Consequently, we find that all but perhaps a few bursts, dark or otherwise, probably occur in molecular clouds.  

In \S 5, we discuss implications for the isotropic-equivalent peak luminosity of the optical flash, the density of the circumburst medium, circumburst extinction studies, and the nature of star formation in the universe:  We find that for most bursts, the isotropic-equivalent peak luminosity of the optical flash is probably less than $L_{49} \sim 0.1$ (\S 5.1), that the density of the circumburst medium is possibly higher than what efforts to model X-ray through radio afterglow data have found to date (\S 5.2), that bursts might serve as probes of dust under the extreme conditions of sublimation and fragmentation, and as a function of redshift (\S 5.3), and that our finding that most, if not all, bursts occur in molecular clouds strengthens the pre-existing idea that most of the star formation in the universe occurs in molecular clouds (\S 5.4).  We summarize our conclusions in \S 6.

\section{Evidence for a Molecular Cloud Origin for Dark Bursts:  Cloud Equilibrium, Gravitational Collapse, and Fragmentation-Driven Bursts of Star Formation}

We first consider the equilibrium properties of diffuse clouds:  i.e., gas that is bound by pressure equilibrium with a warm or hot phase of the interstellar medium.  Specifically, Spitzer (1978) considers the equilibrium properties of a uniformly magnetized, isothermal, non-rotating diffuse cloud, and introduces factors $c_1 = 0.53$ and $c_2 = 0.60$ that correct for the fact that in equilibrium the cloud will not be of uniform hydrogen density $n_H$, but will be centrally condensed, and will not be spherical with radius $R$, but will be moderately flattened along the direction of the magnetic field.  Spitzer (1978) shows that clouds with $R > R_m$ are unstable to gravitational collapse, where
\begin{equation}
\frac{GM^2}{R_m^4} = \frac{25p_m}{1-(M_c/M)^{2/3}},
\end{equation}
$M = 4\pi\rho R_m^3/3$ is the mass of the cloud, $\rho = \mu_Hm_Hn_H$, $\mu_H = 1.87$ is the mean molecular weight per hydrogen atom, 
\begin{equation}
p_m = \frac{3.15c_2(kT/\mu)^4}{G^3M^2[1-(M_c/M)^{2/3}]^3},
\end{equation}
$T$ is the temperature of the cloud, $\mu = 1.44$ is the mean molecular weight, 
\begin{eqnarray}
M_c & = & \frac{0.0236(c_1B)^3}{G^{3/2}\rho^2} \cr
& \approx & 10^4\left(\frac{B}{1\,\rm{\mu G}}\right)^3\left(\frac{n_H}{1\,\rm{cm^{-3}}}\right)^{-2}\,\rm{M_{\sun}},
\end{eqnarray}
and $B$ is the strength of the magnetic field within the cloud.  We plot $R_m(n_H)$ for $T = 10$ and 100 K and $B = 3$ $\mu$G in Figure 1, and for $T = 30$ K and $B = 1$ and 10 $\mu$G in Figure 2:  Most Galactic clouds have $10 < T < 100$ K, and $B = 3$ $\mu$G is typical of the Galactic interstellar medium.

Also in Figures 1 and 2, we plot the curves of constant post-sublimation/fragmentation 1 -- 7.5 eV optical depth $\tau$ from Figures 3 and 4 of Reichart (2001b):  The three pairs of solid curves mark $\tau = 0.3$ (lower left) and 3 (upper right) for $L_{49} = 0.1$ (left), 1 (center), and 10 (right).  For a given $L$, bursts that occur to the lower left of the $\tau = 0.3$ curve have afterglows that are relatively unextinguished by the circumburst cloud, and bursts that occur to the upper right of the $\tau = 3$ curve have highly extinguished afterglows.  Since circumburst extinction appears to be responsible for most of the dark bursts (Reichart \& Yost 2001), we find in Reichart (2001b) that most of the dark bursts occur to the upper right of the $\tau = 3$ curve, which implies relatively large sizes and high densities for their circumburst clouds (\S 1).  
We show in Figures 1 and 2 that diffuse clouds of these sizes and densities are unlikely to be in equilibrium, or even near equilibrium, for typical temperatures and magnetic field strengths.  Such clouds would be unstable to gravitational collapse, resulting in the collapse and fragmentation of the cloud until star formation re-establishes pressure equilibrium within the fragments, and the fragments are bound by self-gravity:  This is precisely the structure of molecular clouds (e.g., Solomon et al. 1987).  Furthermore, molecular clouds regularly have sizes $R \ga 10$ pc and masses $M \ga 3\times10^5$ M$_{\sun}$, and are active regions of star formation (see \S 5.4).  Consequently, dark bursts probably occur in collapsing diffuse clouds/forming molecular clouds, and/or molecular clouds that re-established equilibrium some time ago.  We confirm that molecular clouds have sufficiently high column densities to make a large fraction of bursts dark at optical wavelengths in \S 3.

Solomon et al. (1987) find the mass scale of the molecular cloud fragments to be on the order of a few solar masses, which supports the idea that they are supported against further collapse and fragmentation by star formation.  Since the lifetimes of massive stars are comparable to the free fall time of the progenitor cloud for these cloud masses, the first generation of massive star formation and supernovae should occur on this short timescale:  $\sim 10^7 - 10^8$ yr.  Consequently, dark bursts are probably a byproduct of this burst of star formation if the molecular cloud formed recently, and/or the result of lingering or latter generation star formation if the molecular cloud formed some time ago.  

\section{Consistency Check:  The Column Density and Optical Depth Distributions of Bursts in Galactic-Like Molecular Clouds}

We now consider the column density and optical depth properties of bursts in Galactic-like molecular clouds.\footnote{We must entertain the possibility that Galactic molecular clouds are not representative of molecular clouds elsewhere in the universe, in which case the strength of our conclusions is diminished.  However, since molecular clouds are formed from diffuse clouds that become too small or dense to support their mass, it seems that environmental differences between galaxies more likely affect the numbers of molecular clouds formed than the bulk physical properties of these clouds.  This is supported by the uniformity of molecular cloud mean column densities across of the varied environments of our own galaxy (see below).}  First, we compute the mean column densities of the 273 molecular clouds in the sample of Solomon et al. (1987), which was constructed using data from the Massachusetts-Stony Brook CO Galactic Plane Survey.  The mean column density of a cloud is given by $N_H = M/(\mu_Hm_HA)$, where $M$ is the virial mass (Solomon et al. 1987 show that the clouds are in or near virial equilibrium), $A = 11.56(D\tan\sqrt{\sigma_l\sigma_b})^2$ is the effective cross section of the cloud (Solomon et al. 1987), $D$ is the distance to the cloud, and $\sigma_l$ and $\sigma_b$ are the angular sizes of the cloud.  We plot the distribution (dotted histogram) of mean column densities in Figure 3, and confirm the finding of Solomon et al. (1987) that $\mu_HN_H$ is narrowly peaked around $\approx 170$ M$_{\sun}$ pc$^{-2}$, independent of cloud mass and size.

Next, we correct this distribution for a number of effects.  First, since bursts more likely trace cloud mass than cloud number, we weight these mean column densities by cloud mass, and renormalize the distribution (dashed histogram).  Since molecular cloud column densities appear to be fairly independent of cloud mass, the distribution is not significantly changed.  Also, we weight the $M < 7\times10^4$ M$_{\sun}$ clouds by an additional factor of  $(M/7\times10^4$ M$_{\sun})^{-3/2}$ to correct for an undercounting of low mass clouds on the far side of the Galaxy (Solomon et al. 1987).  However, this affects the mass-weighted mean column density distribution negligibly.

Finally, we correct for geometrical effects:  (1) bursts occur in their circumburst clouds, not behind them; and (2) molecular clouds are centrally condensed.  First, we adopt the same density distribution for the clouds that Solomon et al. (1987) adopted to compute the virial masses:  $\rho \propto r^{-1}$ within the effective radius of the cloud, and $\rho = 0$ beyond this radius.  This density distribution results in a surface brightness profile that is similar to what is observed (Solomon et al. 1987).  Next, we trace this density distribution with $10^4$ randomly-placed bursts for each cloud, and compute for each burst the distance and column density to the earth-facing side of the cloud.  The effect of placing bursts in the clouds, as opposed to behind the clouds, is to lower the mean column densities on average by a factor of two, and sometimes (when bursts occur near the earth-facing side of the cloud) considerably more.  However, this is partially offset by the central condensation of the clouds, which places more bursts in above average density (and column density) environments.  We plot the final distribution -- the expected column density distribution for bursts in Galactic-like molecular clouds -- also in Figure 3 (solid histogram).

Lastly, we confirm that Galactic-like molecular clouds, the column densities of which appear to be fairly universal, at least within the Galaxy, have sufficiently high column densities to make a large fraction of bursts dark at optical wavelengths.  To this end, we plot in Figure 4 the expected distribution (1 and 2 $\sigma$ dashed contours) of column densities and distances to the earth-facing side of the cloud for bursts in Galactic-like molecular clouds.  Also in Figure 4, we plot the curves of constant post-sublimation/fragmentation optical depth from Figures 1 and 2, and lines of constant column density.  It appears that for low values of the isotropic-equivalent peak luminosity of the optical flash ($L_{49} \la 0.1$), Galactic-like molecular clouds can make a large fraction of bursts dark at optical wavelengths.  However, even for very low values of $L$, a reasonable fraction of bursts would be left relatively unextinguished at optical wavelengths.  Since comparable fractions of dark bursts and bursts with detected optical afterglows are observed (about 2/3 and 1/3, respectively; e.g., Fynbo et al. 2001; Lazzati, Covino \& Ghisellini 2001), this suggests that the bursts with detected optical afterglows might also occur in molecular clouds.  We confirm this by modeling and constraining the distribution of column densities, measured from absorption of the X-ray afterglow, of the bursts with detected optical afterglows in \S 4.  We return to the issue of the isotropic-equivalent peak luminosity of the optical flash in \S 5.1.

\section{Evidence for a Molecular Cloud Origin for Bursts with Detected Optical Afterglows:  The Column Density Distribution}

In this section, we model and constrain the column density distribution of the bursts with detected optical afterglows, and compare this distribution to the expected distribution for bursts in Galactic-like molecular clouds (\S 3).  To this end, we have reviewed the literature, and list in Table 1 observer-frame column densities and uncertainties that have been measured from the X-ray afterglows of 15 bursts, 11 of which have detected optical afterglows (this information is not yet available for the vast majority of the dark bursts).  Before modeling these data, we convert each measurement $N_{H,i}$ and upper and lower 1-$\sigma$ uncertainty $\sigma_{u,i}$ and $\sigma_{l,i}$ into probability distributions, first in the observer frame, and then in the source frame by correcting the observer-frame probability distribution for the Galactic column density along the line of sight, and the redshift of the burst if known.  If $\sigma_{u,i} = \sigma_{l,i}$ (or $\sigma_{l,i} = N_{H,i}$), we take the observer-frame probability distribution to be a Gaussian of mean $N_{H,i}$ and standard deviation $\sigma_{u,i}$:  $p_i(N_{H,obs}) = G(N_{H,obs},N_{H,i},\sigma_{u,i})$.  Otherwise, we take the observer-frame probability distribution to be a skewed Gaussian:  $p_i(N_{H,obs}) = G(N_{H,obs}^s,N_{H,i}^s,\sigma_i^s)dN_{H,obs}^s/dN_{H,obs}$, where $s$ is the skewness parameter, and $s$ and $\sigma_i$ are given by solving 
\begin{eqnarray}
\sigma_i^s & = & (N_{H,i}+\sigma_{u,i})^s-N_{H,i}^s \cr
& = & N_{H,i}^s-(N_{H,i}-\sigma_{l,i})^s,
\end{eqnarray}
i.e., we take the observer-frame probability distribution to be a Gaussian in $N_{H,obs}^s$, which reduces to a Gaussian in $N_{H,obs}$ when $\sigma_{u,i} = \sigma_{l,i}$ ($s = 1$).  The source-frame probability distribution $p_i(N_H)$ is then given by substituting $N_{H,obs} = N_{H,MW}+N_H(1+z)^{-2.6}$, where $N_{H,MW}$ is the Galactic column density along the line of sight (interpolated\footnote{We use the $n_H$ FTOOL at http://heasarc.gsfc.nasa.gov/cgi-bin/Tools/w3nh/w3nh.pl.} from the maps of Dickey \& Lockman 1990), $N_H$ is the source-frame column density, and $(1+z)^{-2.6}$ scales $N_H$ to the observer frame (e.g., Morrison \& McCammon 1983).  If multiple column density measurements are available for a burst, we average their source-frame probability distributions if the measurements were made from the same data, and we take the product of these distributions if the measurements were made from independent data (GRB 000926).  We plot the source-frame probability distributions, peak-normalized to one, for the bursts with detected optical afterglows in the top panel of Figure 5.  The dotted curves mark bursts of unknown redshift, in which case we fix $z = 0$:  These distributions can be thought of as fuzzy lower limits (see below).  

We model the column density distribution of the bursts with detected optical afterglows with a broken power law:
\begin{equation}
n(\log{N_H}) = \cases{0 & ($N_H < 0$) \cr 
\frac{1}{\ln{10}}\frac{bc}{c-b}\left(\frac{N_H}{a}\right)^b & ($0 < N_H < a$) \cr 
\frac{1}{\ln{10}}\frac{bc}{c-b}\left(\frac{N_H}{a}\right)^c & ($N_H > a$)},
\label{model}
\end{equation}
where $a$ is the break column density, $b$ is the low column density power-law index, and $c$ is the high column density power-law index.  We use a three-parameter function for flexibility, and a simple function (as opposed to, e.g., the more elegant skewed Gaussian) so the inner integral of Equation (\ref{double}) can be evaluated analytically (otherwise, the error analysis would be computationally taxing).  We note that Equation (\ref{model}) is log-normalized [$\int_{-\infty}^{\infty}n(\log{N_H})d\log{N_H} = 1$] to facilitate comparison of the fitted distribution to the solid histogram of Figure 3 (see below).

We fit the model $n(\log{N_H})$ to the data $p_i(N_H)$ using Bayesian inference (e.g., Reichart 2001a).  The likelihood function is given by
\begin{equation}
{\cal L} = \prod_{i=1}^{N}{\cal L}_i,
\end{equation}
where $N = 11$ is the number of bursts with detected optical afterglows in our sample, and
\begin{equation}
{\cal L}_i = \cases{\int_{-\infty}^{\infty}p_i(N_H)n(N_H)dN_H & ($z_i$ known) \cr
\int_{-\infty}^{\infty}p_i(N_H')\left[\int_{N_H'}^{\infty}n(N_H)dN_H\right]dN_H' & ($z_i$ unknown)} 
\end{equation}
(e.g., Reichart \& Yost 2001), or equivalently
\begin{equation}
{\cal L}_i = \cases{\int_{-\infty}^{\infty}p_i(N_H)n(\log{N_H})d\log{N_H} & ($z_i$ known) \cr
\int_0^{\infty}p_i(N_H')\left[\int_{\log{N_H'}}^{\infty}n(\log{N_H})d\log{N_H}\right]dN_H' & ($z_i$ unknown)},
\label{double}
\end{equation} 
since $n(\log{N_H}) = 0$ for $N_H < 0$.  We find that $\log{(a/{\rm cm}^{-2})} = 22.16^{+0.20}_{-0.33}$, arctan$\,b = 0.92^{+0.38}_{-0.35}$ ($b \approx 1.31^{+2.25}_{-0.68}$), and arctan$\,c = -1.46^{+0.53}_{-0.11}$ [$c \la -1.33$ (1 $\sigma$)].  We plot the likelihood-weighted average model distribution (solid curve), which is also log-normalized, and the 1-$\sigma$ uncertainty in this distribution (dotted curves) in the bottom panel of Figure 5.  We find that the column density distribution of the bursts with detected optical afterglows peaks around $10^{22}$ cm$^{-2}$, and spans a factor of a few to an order of magnitude or so.  These findings verify and improve upon the earlier finding of Galama \& Wijers (2001) that these column densities span $10^{22} - 10^{23}$ cm$^{-2}$.

Lastly, we compare the fitted column density distribution of the bursts with detected optical afterglows to the expected column density distribution for bursts in Galactic-like molecular clouds (solid histogram, replotted from Figure 3), and find these distributions to be consistent within the 1-$\sigma$ uncertainties.  Furthermore, the fitted distribution is not consistent with the expectation for bursts that occur in diffuse clouds, the canonical column density of which is a few times $10^{20}$ cm$^{-2}$.  Consequently, we find that all but perhaps a few bursts, dark and otherwise, probably occur in molecular clouds.

\section{Discussion}

\subsection{Implications for the Isotropic-Equivalent Peak Luminosity of the Optical Flash}

Now that we have shown that all but perhaps a few bursts probably occur in molecular clouds (\S 4), we return to the issue of the isotropic-equivalent peak luminosity of the optical flash.  In \S 3, we showed that Galactic-like molecular clouds can make a large fraction of bursts dark at optical wavelengths, but only if a large fraction of bursts have $L_{49} \la 0.1$.  Furthermore, even if the value of $L$ were very low, a reasonable fraction of these bursts would be sufficiently near the earth-facing side of their circumburst clouds, and/or in sufficiently small clouds, to be left relatively unextinguished at optical wavelengths (Figure 4).  Since comparable fractions of dark bursts and bursts with detected optical afterglows are observed, this suggests that most bursts, dark or otherwise, have $L_{49} \la 0.1$, and that bursts with $L_{49} \ga 1$ should be relatively rare.  Indeed, only one such burst, GRB 990123 (Akerlof et al. 1999) with $L_{49} \approx 1$ (Waxman \& Draine 2000), has been identified to date.  Furthermore, of 11 other bursts that were observed on this timescale by ROTSE I, all have flux upper limits that are fainter than the peak flux of the optical flash of GRB 990123, often by 4 -- 5 magnitudes (Akerlof et al. 2000; Kehoe et al. 2001).  Similar results have been found with LOTIS (e.g., Williams et al. 2000). 

These findings are consistent with the simple model introduced by Reichart \& Yost (2001), in which the collimation angle of a burst's ejecta, which strongly affects the value of $L$, and the column density to the earth-facing side of the circumburst cloud, determine whether a burst is dark or not.  

\subsection{Implications for the Density of the Circumburst Medium}

Figure 4 suggests that bursts typically occur in regions of density $n_H \approx 30 - 10^3$ cm$^{-3}$.  However, to date, the modeling of X-ray through radio afterglow data has suggested significantly lower densities, ranging from $n \sim 5\times10^{-4}$ cm$^{-3}$ (GRB 990123; Panaitescu \& Kumar 2001) to $n \approx 30$ cm$^{-3}$ (GRB 000926; Harrison et al. 2001).  A number of reasonable solutions to this problem come to mind.  First, the densities in Figure 4 are average densities to the earth-facing side of the circumburst cloud, but molecular clouds consist of fragments bound by self-gravity (\S 2).  Solomon et al. (1987) estimate that there are $\approx 10$ fragments along a typical line of sight through a typical molecular cloud, that the density of these fragments is on the order of $10^3$ cm$^{-3}$, and that the size of these fragments is on the order of a few tenths of a parsec, which is very roughly the distance a burst's ejecta travel before the burst occurs.  Consequently, low densities might be expected if the ejecta clear the circumburst fragment before the burst occurs, or at least before the observed afterglow occurs.\footnote{If the ejecta clear the circumburst fragment after the burst occurs, one might be able to measure a change in the X-ray column density.  Although this would be only a 10 -- 20\% effect for most bursts (assuming typically 5 -- 10 fragments between the burst and the earth-facing side of the cloud), a few bursts should occur sufficiently close to the earth-facing side of the cloud to make this a 50 -- 100\% effect.}  Alternatively, the circumburst fragment might be dispersed by prior massive star formation within the fragment (e.g., Scalo \& Wheeler 2001).  

Another reasonable solution to this problem is that the practice of modeling afterglow data is not yet, or is just now becoming, sufficiently sophisticated to correctly extract the density of the circumburst medium.  Case in point is that only very recently has the inverse-Compton component of the afterglow been included in modeling efforts:  This component is essential to correctly interpret afterglow data if $n \ga 1$ cm$^{-3}$ (e.g., Sari \& Esin 2001).  Including this component in their model, Harrison et al. (2001) found the first observational evidence for this component in the afterglow of GRB 000926:  The strength of this component implied a density for the circumburst medium of $n \approx 30$ cm$^{-3}$, the highest value yet reported.  However, more interesting is the fact that the source-frame column density of this burst is the most stringently low of all of the bursts (of known redshift) in the top panel of Figure 5.  This suggests that even higher densities might be found for the circumburst environments of the other bursts with detected optical afterglows once properly modeled, or once this region of the parameter space is more fully explored.

Lastly, we point out that these modeling efforts are naturally biased against the identification of high density circumburst environments.  The self-absorption frequency increases with increasing density of the circumburst medium (e.g., Sari, Piran \& Narayan 1998; Sari \& Esin 2001):  Once $n \ga 10 - 10^2$ cm$^{-3}$ (for typical values of other parameters; Reichart \& Yost 2001), the radio afterglow becomes increasingly difficult to detect.  Also, if a few parsecs of dust remain along the line of sight, once $n \ga 10^2$ cm$^{-3}$, the optical afterglow becomes increasing difficult to detect.  Since modelers tend to avoid radio dark and/or optically dark bursts, they naturally bias themselves against the identification of circumburst environments of these densities.

We address this problem in a simple way in Reichart \& Yost (2001):  We show that a range of densities for the circumburst medium, with other parameters fixed to canonical values, can explain snapshot optical vs. X-ray and optical vs. radio brightness distributions of the afterglows of all bursts, radio dark and optically dark bursts included.  We find that the density of the circumburst medium probably spans many orders of magnitude, from densities possibly as low as densities that are typical of the Galactic disk to densities probably as high as densities that are typical of dense clouds.  

\subsection{Implications for Circumburst Extinction Studies}

Figure 4 also suggests that if most bursts have $L_{49} \la 1$, which appears to be the case (\S 5.1), a reasonable fraction of bursts should have afterglows that are extinguished at intermediate levels:  $0.3 \la \tau \la 3$, where $\tau \sim$ source-frame $A_{\rm B}$ (Reichart 2001b).  Such bursts would be the missing link between dark bursts and bursts with relatively unextinguished afterglows.  Furthermore, by extracting extinction curves from the afterglow data of such bursts (Reichart 2001a), bursts could serve as probes of dust under the extreme conditions of sublimation and fragmentation, and as probes of dust as a function of redshift, to very high redshifts (Lamb \& Reichart 2000).

\subsection{Implications for the Nature of Star Formation in the Universe}

Mooney \& Solomon (1988) find that the far-infrared luminosity-to-mass ratio for isolated and weakly interacting spiral galaxies is consistent with the average for Galactic molecular clouds, which suggests that most, if not all, of the star formation in such galaxies occurs in molecular clouds.  Our finding that most, if not all, bursts occur in molecular clouds would strengthen this idea if most of the star formation in the universe occurs in such galaxies (see Lamb \& Reichart 2000 for a review of evidence linking bursts to star formation and massive star death).

However, Ramirez-Ruiz, Trentham \& Blain (2001) argue that about 3/4 of the star formation in the universe occurs in ultraluminous infrared/submillimeter-bright galaxies.  Often, this emission is concentrated within hundreds of parsecs of the center of the galaxy.  Since the column densities to stars in these nuclear regions range from $N_H \approx 10^{23} - 10^{24}$ cm$^{-2}$, and the sizes of these regions are one to two orders of magnitude too large for dust sublimation and fragmentation to matter, bursts that occur in these regions would necessarily be dark at optical wavelengths.  Since about 2/3 of bursts are dark at optical wavelengths, Ramirez-Ruiz, Trentham \& Blain (2001) argue that most of the dark bursts probably occur in such regions.

We now test this idea.  In Figure 6, we plot the source-frame column density probability distributions of the four dark bursts in Table 1 (\S 4).  Redshifts are not known for three of these bursts, so we adopt $z = 1$ (top panel), 2 (middle panel), and 3 (bottom panel) for these three bursts.  Given the redshifts that have been measured for bursts to date, the probabilities that all three of these bursts have $z > 1$, 2, and 3 are 0.27, 0.013, and 0.0016, respectively.  Even in the very unlikely event that all three of these bursts have $z \approx 3$, these four probability distributions are not consistent with the expected column density distribution for bursts that occur in the nuclear regions of ultraluminous infrared/submillimeter-bright galaxies, but are consistent with the expected column density distribution for bursts that occur in molecular clouds (\S\S 3 and 4).

However, these ideas can be reconciled if the stars, and in particular the massive stars, form predominately outside of the nuclear region.  Ultraluminous infrared/submillimeter-bright galaxies are thought to be strongly interacting spiral galaxies, where the interaction causes a starburst (e.g., Solomon \& Sage 1988; Solomon et al. 1997).  The light from the starburst is absorbed by dust in the star-forming regions, and is re-emitted in the far infrared.  We now consider how the interaction causes the starburst, and where the stars are when this occurs.  We consider first the early stages of the interaction.  Solomon et al. (1997) argue that the starburst cannot be caused by the collisions of molecular clouds, or by the creation of new molecular clouds from the collisions of diffuse clouds, since the interaction velocity exceeds the escape velocity of even the most massive clouds by more than an order of magnitude.  Furthermore, they argue that molecular cloud collisions are improbable, since their mean free path exceeds typical disk thicknesses by about an order of magnitude.  However, diffuse cloud collisions are very probable, the result of which should be a hot, ionized gas that surrounds the molecular clouds in that region of the galaxy (Jog \& Solomon 1992).  This high-pressure gas should cause radiative shock compression of the outer layers of these molecular clouds, resulting in a starburst for which Jog \& Solomon (1992) estimate a far-infrared luminosity of a few times $10^{11}$ L$_{\sun}$.  Starbursts of this nature would span large regions of the galaxy, and indeed, starbursts in colliding galaxies are observed to span several kiloparsecs (e.g., Wright et al. 1984; Joseph \& Wright 1985; Rieke et al. 1985; Wright et al. 1988).  Such starbursts are consistent with Figure 6.  

We now consider the latter stages of the interaction.  Tidal interaction is thought to create a dense nuclear region:  The perturbation of cloud orbits increases the diffuse cloud collision rate, which results in a dissipation of kinetic energy that causes most of the gas in the disk to fall to the center of the galaxy (e.g., Noguchi \& Ishibashi 1986; Hernquist 1989).  However, dense gas alone does not a starburst make.  Jog \& Das (1992) propose that nuclear starbursts are caused by molecular clouds that tumble into the nuclear region:  Again, high-pressure gas causes radiative shock compression of the outer layers of the molecular clouds, resulting in a starburst for which Jog \& Das (1992) estimate a far-infrared luminosity that ranges from $\ga 10^9$ L$_{\sun}$ for distant tidal interactions to $\la 10^{12}$ L$_{\sun}$ for very perturbed mergers.  Even nuclear starbursts might be consistent with Figure 6 if the molecular clouds are shock compressed $\sim 10^7 - 10^8$ yr before they fall into the very high column density environment of the inner nuclear region:  This would give them sufficient time to form and expend their massive stars in low and/or intermediate column density environments.

Alternatively, the fraction of star formation in the universe that occurs in ultraluminous infrared/submillimeter-bright galaxies might be smaller than $\sim 3/4$.  This would imply that (1) the IMF is high-massed biased in SCUBA galaxies, (2) SCUBA sources are powered predominately by AGN, and/or (3) the star-formation rates in UV-bright galaxies are underestimated (Ramirez-Ruiz, Trentham \& Blain 2001).

However, the idea that bursts should be associated with ultraluminous infrared/submillimeter-bright galaxies is supported by the recent associations of GRB 980703 with an ultraluminous infrared galaxy (Berger, Kulkarni \& Frail 2001), and GRB 010222 with a submillimeter-bright galaxy (Frail et al. 2001).  However, since both GRB 980703 and GRB 010222 have well-detected optical afterglows, they could not have occurred in dense nuclear regions, and hence are in agreement with Figure 6 and the above possible explanations.

In any case, our finding that most, if not all, bursts occur in molecular clouds strengthens the idea that most of the star formation in the universe occurs in molecular clouds, be they in isolated or interacting galaxies.

\section{Conclusions}

We find that the circumburst clouds of dark bursts are too large and massive to be diffuse, and consequently are probably molecular, making them active regions of star formation.  Furthermore, we show that Galactic molecular clouds, the column densities of which appear to be fairly universal, at least within the Galaxy, have sufficiently high column densities to make a large fraction of bursts dark at optical wavelengths.  However, since the optical flash sublimates dust and the burst and afterglow fragment dust, we find that most bursts in Galactic-like molecular clouds would have to have $L_{49} \la 0.1$:  This seems reasonable, given the relatively deep imaging of ROTSE I and LOTIS.  Furthermore, we find that even for very low values of $L$, a reasonable fraction of bursts in Galactic-like molecular clouds would be sufficiently near the earth-facing side of the cloud, and/or in sufficiently small clouds, to be left relatively unextinguished at optical wavelengths.  Consequently, we model and constrain the distribution of column densities, measured from absorption of the X-ray afterglow, of the bursts with detected optical afterglows:  We find that these bursts are also consistent with a molecular cloud origin, and are not consistent with a diffuse cloud origin.  Consequently, we find that all but perhaps a few bursts, dark or otherwise, probably occur in molecular clouds:  This seems reasonable, given that it is thought that most of the star formation in the universe occurs in molecular clouds, be they in isolated or interacting galaxies.
Finally, we show that the limited information that is available on the column densities of the dark bursts is not consistent with the idea that the dark bursts occur in the nuclear regions of ultraluminous infrared/submillimeter-bright galaxies.  However, this does not necessarily imply that most of the star formation in the universe does not occur in ultraluminous infrared/submillimeter-bright galaxies.

\acknowledgements

Support for this work was provided by NASA through the Hubble Fellowship grant \#HST-SF-01133.01-A from the Space Telescope Science Institute, which is operated by the Association of Universities for Research in Astronomy, Inc., under NASA contract NAS5-26555.  P. A. P. acknowledges support from the Alex Rodgers Travelling Scholarship.  We are also grateful to Don Lamb for discussions about the role of magnetic fields in the equilibrium of diffuse clouds, and about the structure of molecular clouds, and to Kara Reichart for helping to input data from Table 1 of Solomon et al. (1987).

\clearpage

\clearpage

\begin{deluxetable}{cccccc}
\tablecolumns{6}
\tablewidth{0pc}
\tablecaption{Column Density Data}
\tablehead{\colhead{GRB} & \colhead{$z$} & \colhead{Dark} & \colhead{$N_{H,MW}$\tablenotemark{a,b}} & \colhead{$N_{H,obs}$\tablenotemark{a}} & \colhead{Ref\tablenotemark{c}}}
\startdata
970228 & 0.695 & N & 1.60 & $3.5^{+3.3}_{-2.3}$ & 1 \nl
& & & & $2.5^{+1.4}_{-1.1}$ & 2 \nl
970402 & & Y & 2.04 & $0.65^{+3.0}_{-0.65}$ & 2 \nl
970508 & 0.835 & N & 0.516 & $1.8^{+2.1}_{-1.1}$ & 2 \nl
970828 & 0.958 & Y & 0.356 & $3.5^{+2.2}_{-1.6}$ & 2 \nl
& & & & $4.1^{+2.1}_{-1.6}$ & 3 \nl
971214 & 3.418 & N & 0.162 & $4.3^{+6.1}_{-2.5}$ & 2 \nl
& & & & $1^{+2.3}_{-1}$ & 4 \nl
980329 & & N & 0.941 & $10\pm4$ & 5 \nl
& & & & $10.7^{+7.8}_{-4.3}$ & 2 \nl
980519 & & N & 1.74 & $5.6^{+10.2}_{-5.1}$ & 2 \nl
990703 & 0.966 & N & 0.580 & $6.6^{+3.8}_{-2.2}$\tablenotemark{d} & 6 \nl
990123 & 1.600 & N & 0.212 & $1.5^{+0.7}_{-0.6}$ & 7 \nl
990510 & 1.619 & N & 0.937 & $2.1\pm0.6$ & 8 \nl
& & & & $1.7\pm0.7$ & 8 \nl
& & & & $1.5\pm0.6$ & 8 \nl
991014 & & Y & 2.45 & $<8.6$ & 9 \nl
991216 & 1.020 & N & 2.04 & $3.5\pm1.5$ & 10 \nl
000214 & & Y & 0.577 & $0.07^{+0.75}_{-0.07}$ & 11 \nl
000926 & 2.037 & N & 0.266 & $0.5\pm0.3$ & 12 \nl
& & & & $0.48\pm0.3$ & 13 \nl
& & & & $0.30\pm0.25$\tablenotemark{e} & 13 \nl
010222 & 1.477 & N & 0.162 & $1.5\pm0.3$ & 14 \nl
\enddata
\tablenotetext{a}{$10^{21}$ cm$^{-2}$.}
\tablenotetext{b}{Interpolated from the maps of Dickey \& Lockman 1990.}
\tablenotetext{c}{1. Frontera et al. 1998; 2. Owens et al. 1998; 3. Yoshida et al. 1999; 4. Dal Fiume et al. 2000; 5. in 't Zand et al. 1998; 6. Vreeswijk et al. 1999; 7. Yonetoku et al. 2000; 8. Kuulkers et al. 2000; 9. in 't Zand et al. 2000; 10. Piro et al. 2000; 11. Antonelli et al. 2000; 12. Piro et al. 2001; 13. Harrison et al. 2001; 14. in 't Zand et al. 2001.}
\tablenotetext{d}{Determined from information in Vreeswijk et al. 1999.}
\tablenotetext{e}{Measurement from independent data.}
\end{deluxetable}

\clearpage

\figcaption[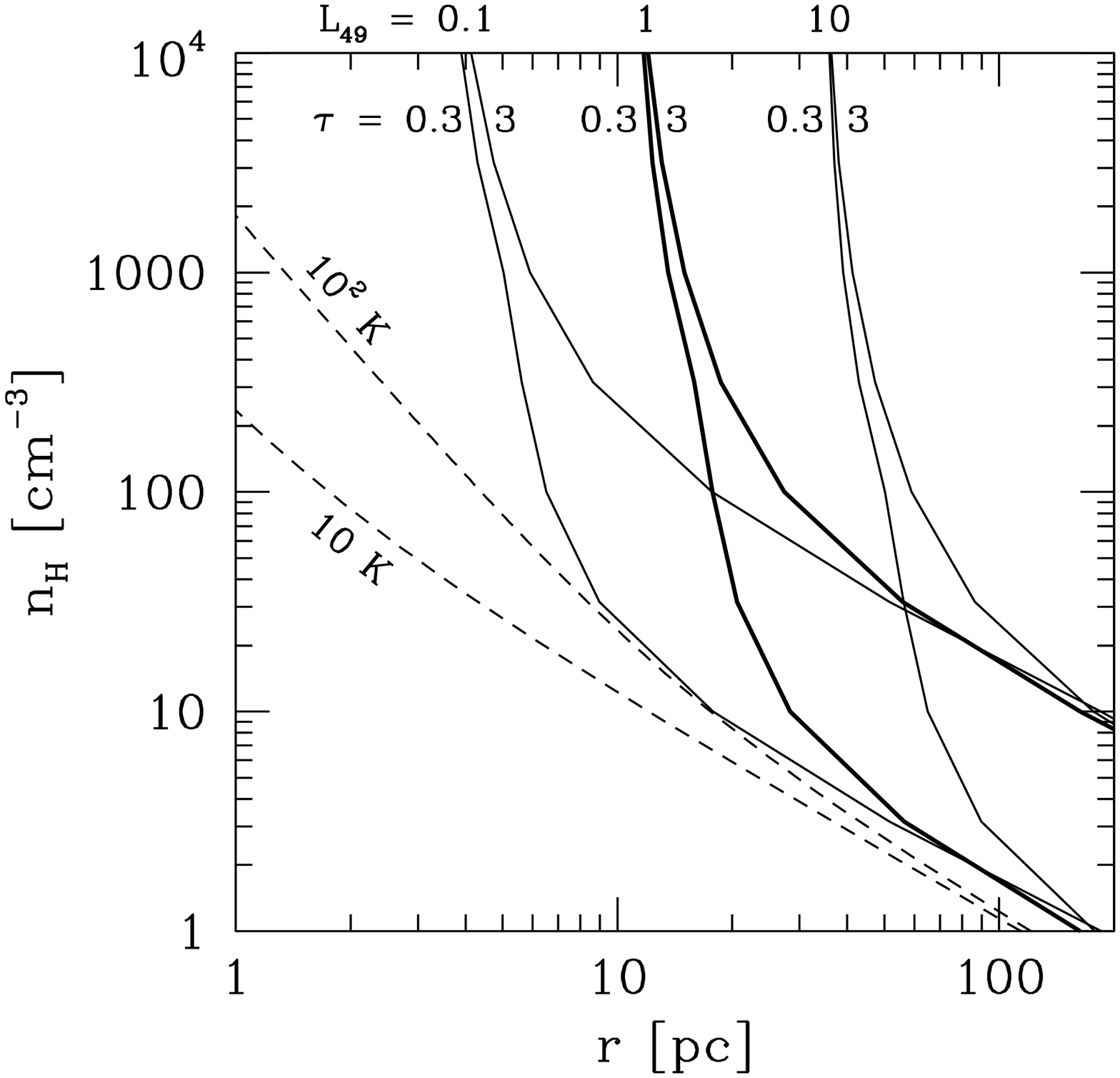]{Post-sublimation/fragmentation optical depth $\tau$ to a burst that is embedded a distance $r$ within a cloud of constant hydrogen density $n_H$, along the line of sight (solid curves; from Reichart 2001b), and diffuse cloud stability criteria (dashed curves).  The three pairs of solid curves mark $\tau = 0.3$ (lower left) and 3 (upper right) for $L_{49} = 0.1$ (left), 1 (center), and 10 (right).  For a given $L$, bursts that occur to the lower left of the $\tau = 0.3$ curve have afterglows that are relatively unextinguished by the circumburst cloud, and bursts that occur to the upper right of the $\tau = 3$ curve have highly extinguished afterglows.  The dashed curves mark the maximum radius for which a diffuse cloud is stable against gravitational collapse, for cloud temperatures $T = 10$ and 100 K and cloud magnetic field strength $B = 3$ $\mu$G (\S 2).\label{taustabtemp.eps}}

\figcaption[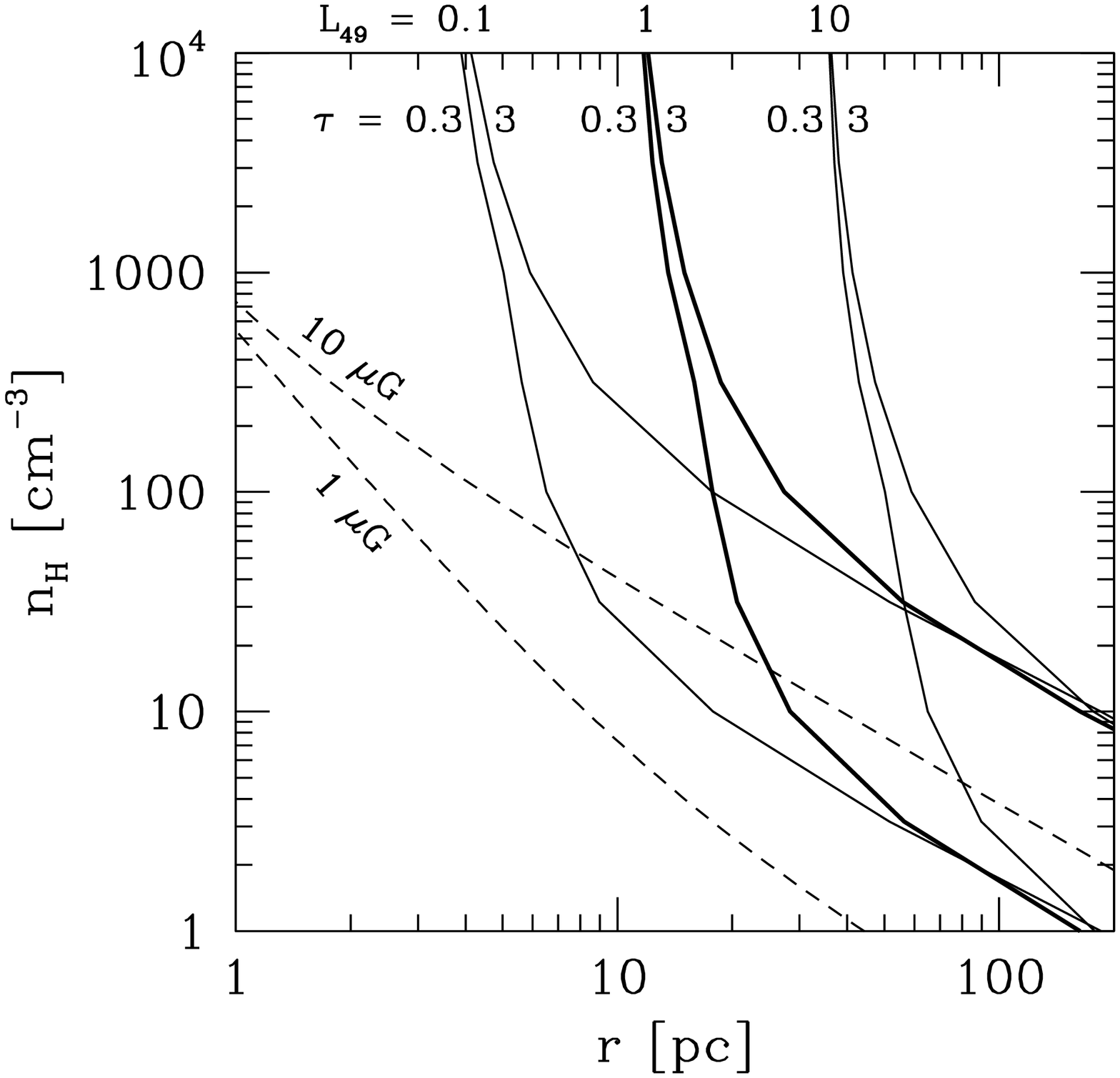]{Same as Figure 1, except for $T = 30$ K and $B = 1$ and 10 $\mu$G (\S 2).\label{taustabmag.eps}}

\figcaption[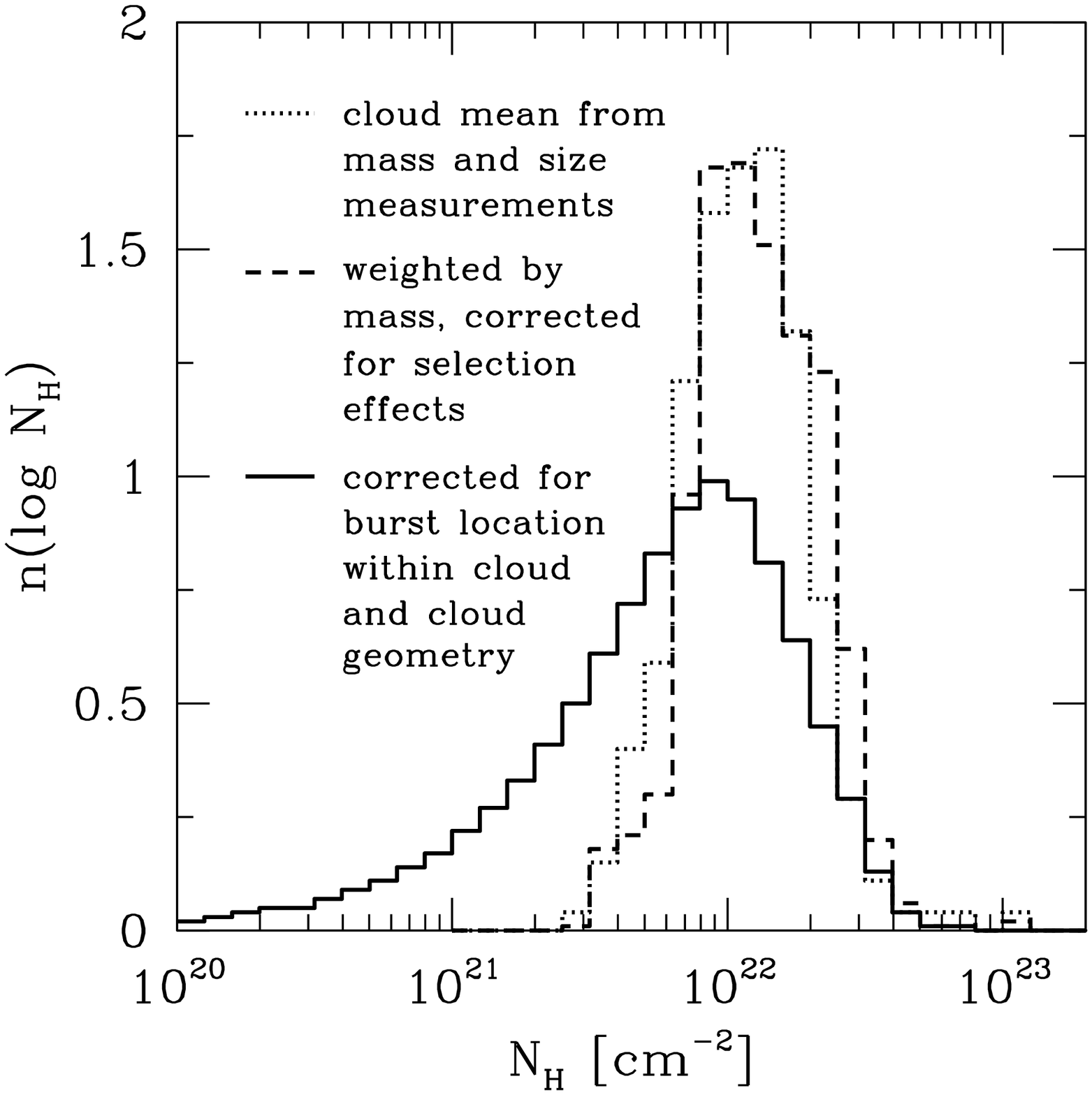]{Expected column density distribution $n(\log{N_H})$ (log-normalized to one) for bursts that occur in Galactic-like molecular clouds (solid histogram).  The dotted histogram marks the mean column density distribution of the Galactic molecular clouds of Solomon et al. (1987), which we compute from their mass and size measurements.  The dashed histogram marks the mass-weighted mean column density distribution, which we also correct for an undercounting of low mass clouds on the far side of the Galaxy (a negligible correction).  The solid histogram marks this distribution after correcting for the fact that bursts occur in their circumburst clouds, not behind them, and for the fact that molecular clouds are centrally condensed (\S 3).\label{nh1.eps}}

\figcaption[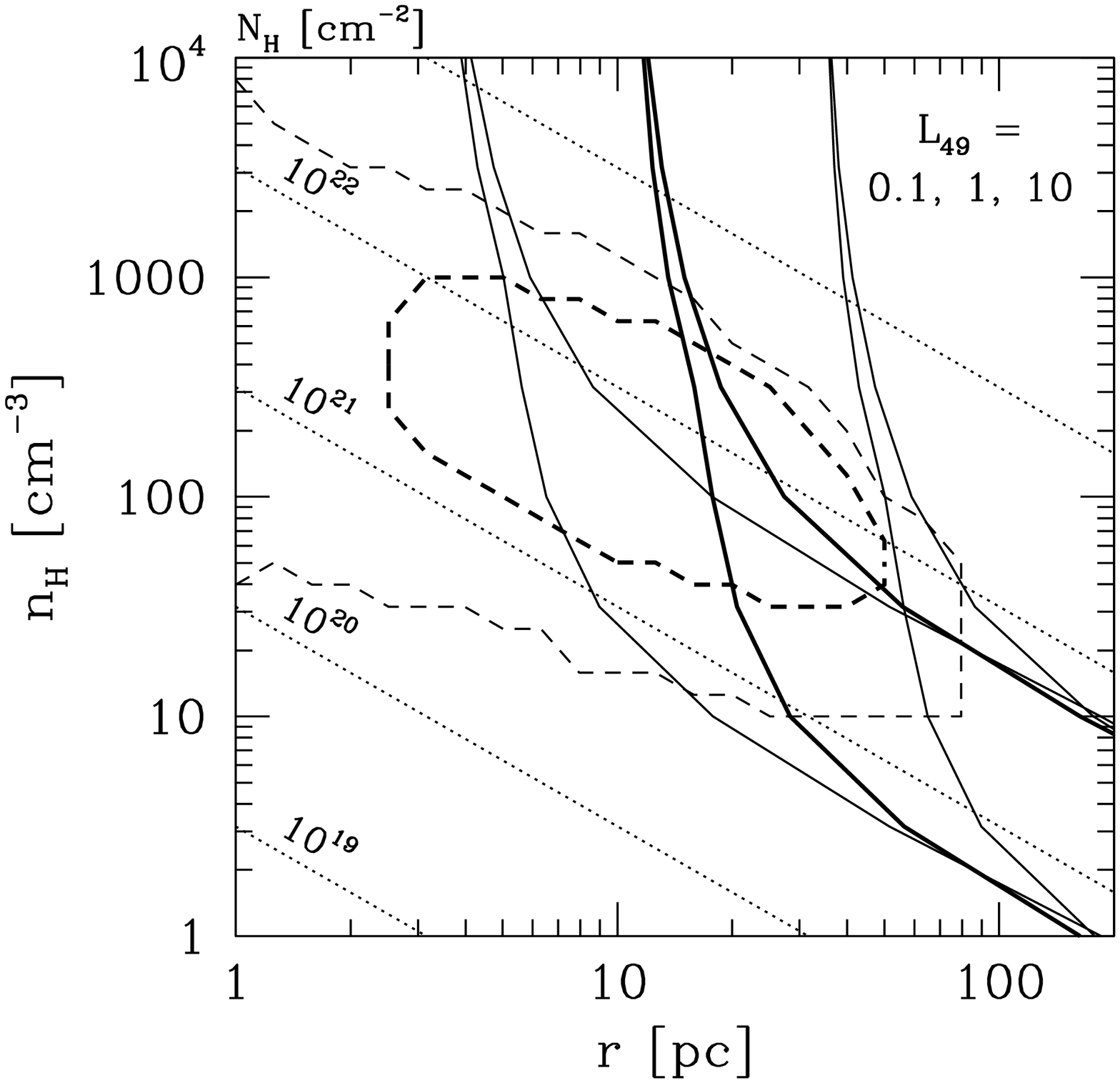]{Expected distribution of column densities and distances to the earth-facing side of the cloud for bursts that occur in Galactic-like molecular clouds (dashed contours), and the curves of constant post-sublimation/fragmentation optical depth from Figures 1 and 2 (solid curves).  The thick dashed contour contains 68.3\% of the distribution, and the thin dashed contour contains 95.4\% of the distribution.  The dotted lines mark constant hydrogen column densities (\S 3).\label{taumc.eps}}

\figcaption[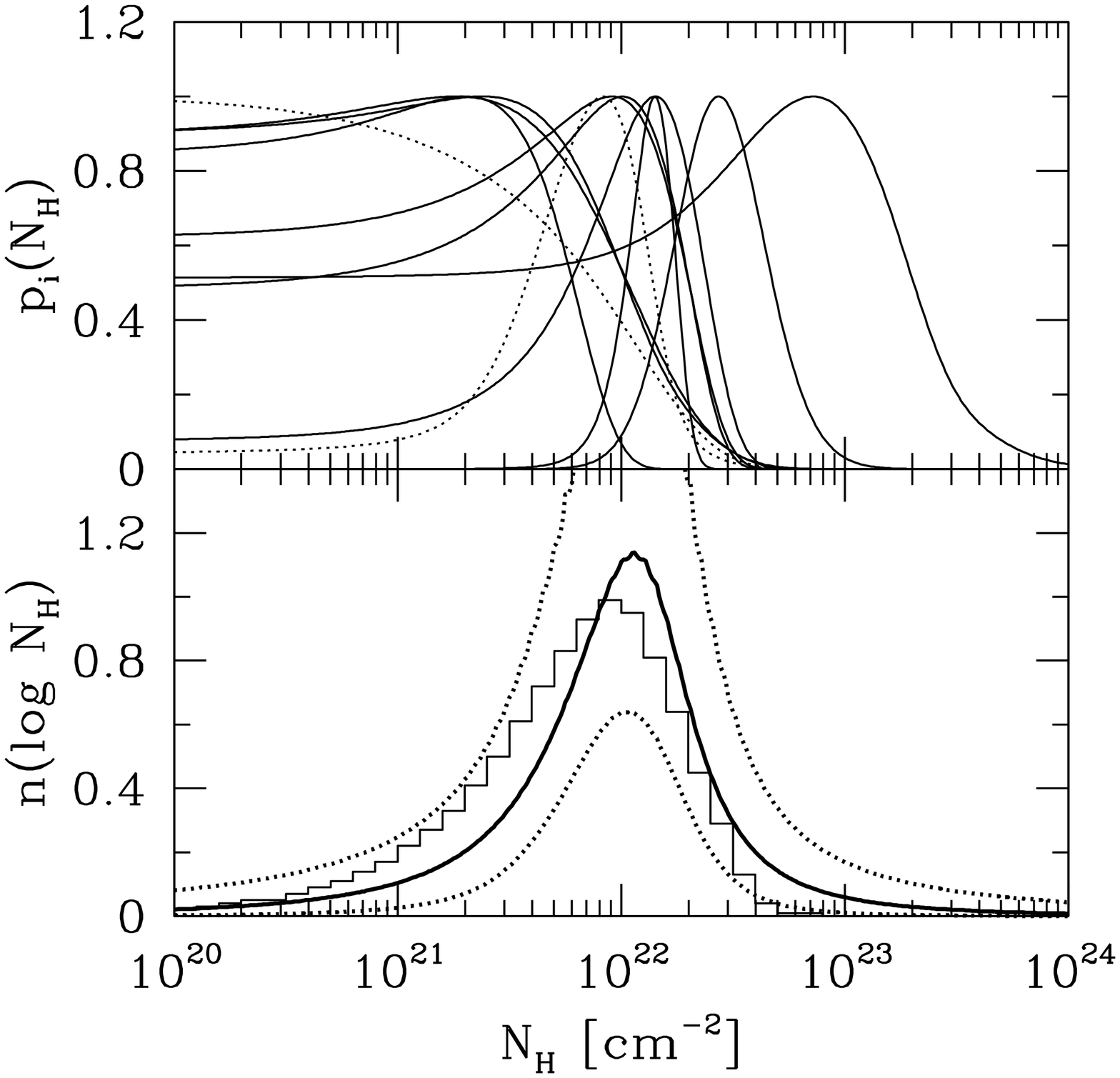]{Top panel:  Source-frame column density probability distributions $p_i(N_H)$ (peak-normalized to one) for the 11 bursts with detected optical afterglows in Table 1.  The dotted curves mark bursts of unknown redshift, for which we fix $z = 0$:  These distributions can be thought of as fuzzy lower limits.  Bottom panel:  Likelihood-weighted average model distribution $n(\log{N_H})$ (solid curve), and 1-$\sigma$ uncertainty in this distribution (dotted curves).  The solid histogram is the expected column density distribution for bursts that occur in Galactic-like molecular clouds from Figure 3.  These distributions are consistent within the 1-$\sigma$ uncertainties (\S 4).\label{nh2.eps}}

\figcaption[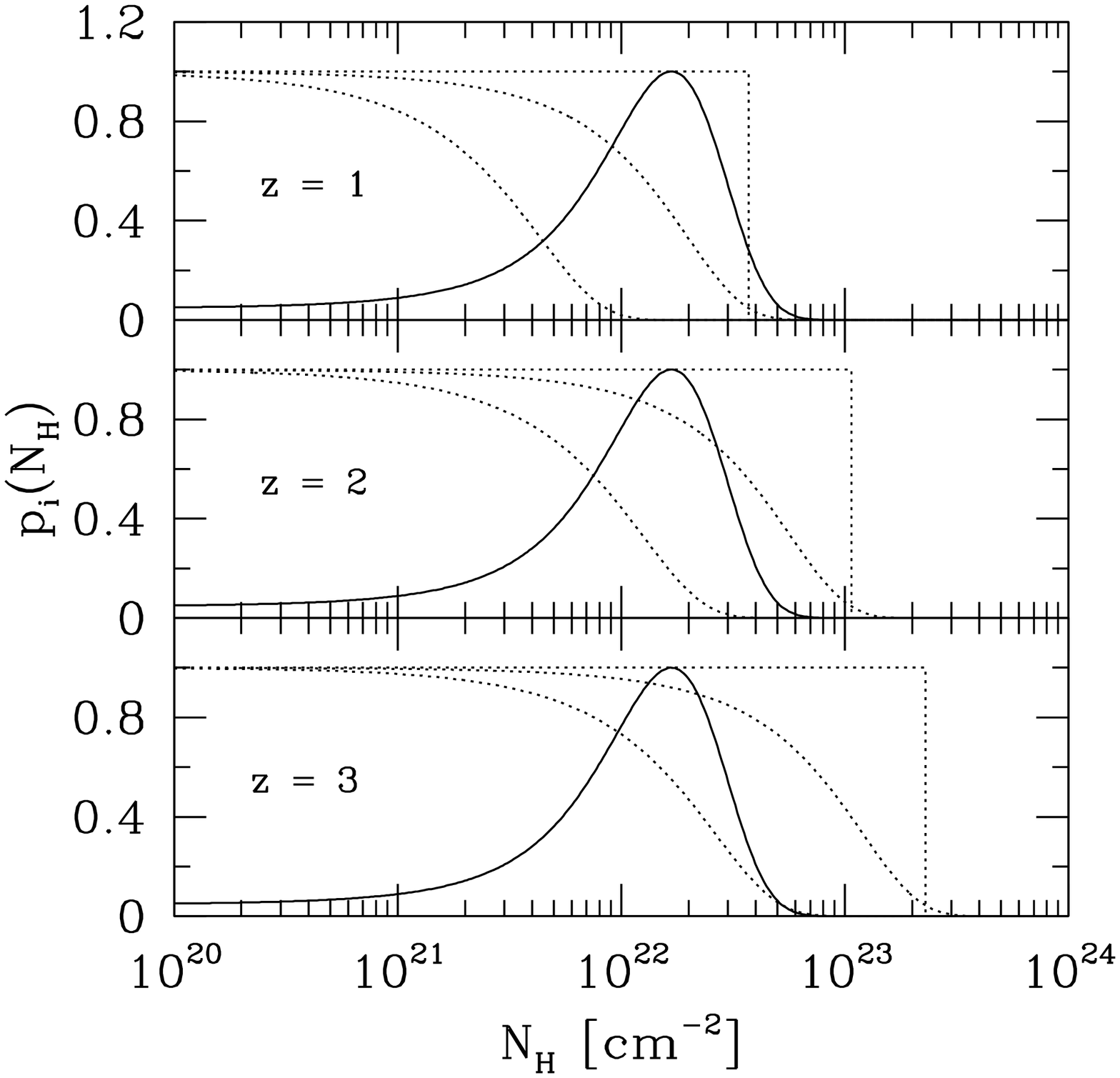]{Source-frame column density probability distributions $p_i(N_H)$ for the four dark bursts in Table 1.  The dotted curves mark bursts of unknown redshift, for which we adopt $z = 1$ (top panel), 2 (middle panel), and 3 (bottom panel; \S 5.4).\label{nh3.eps}}

\clearpage

\setcounter{figure}{0}

\begin{figure}[tb]
\plotone{taustabtemp.eps}
\end{figure}

\begin{figure}[tb]
\plotone{taustabmag.eps}
\end{figure}

\begin{figure}[tb]
\plotone{nh1.eps}
\end{figure}

\begin{figure}[tb]
\plotone{taumc.eps}
\end{figure}

\begin{figure}[tb]
\plotone{nh2.eps}
\end{figure}

\begin{figure}[tb]
\plotone{nh3.eps}
\end{figure}

\end{document}